\documentclass[
reprint,
superscriptaddress,
showpacs,
 amsmath,amssymb,
 aps,
longbibliography,
dvipsnames,
beamer,
floatfix,
]{revtex4-2}

\usepackage{graphicx}
\usepackage{dcolumn}
\usepackage{bm}
\usepackage{amsmath}

\usepackage{textcomp}
\usepackage{siunitx}

\usepackage{xcolor}
\usepackage{ulem}

\begin{document}


\title[]{Microelectronic readout of a diamond quantum sensor}


\author{D. Wirtitsch}
\email{Equal contribution.}
\affiliation{Vienna centre for Quantum Science and Technology, Department of Physics, University of Vienna, Boltzmanngasse 5, 1090 Vienna, Austria}
\affiliation{Austrian Academy of Sciences, Institute for Quantum Optics and Quantum Information (IQOQI) Vienna, Boltzmanngasse 3, A-1090 Vienna, Austria}
\author{G. Wachter}
\email{Equal contribution.}
\affiliation{Vienna centre for Quantum Science and Technology, Department of Physics, University of Vienna, Boltzmanngasse 5, 1090 Vienna, Austria}
\author{S. Reisenbauer}
\affiliation{Vienna centre for Quantum Science and Technology, Department of Physics, University of Vienna, Boltzmanngasse 5, 1090 Vienna, Austria}
\affiliation{AIT Austrian Institute of Technology GmbH, Giefinggasse 4, 1210 Wien}
\author{J. Schalko}
\author{U. Schmid}
\affiliation{ Institute of Sensor and Actuator Systems, TU Wien, Gußhaußstr. 27-29, 1040 Vienna,
Austria}
\author{A. Fant}
\affiliation{Infineon Technologies, Villach, Austria.}
\author{L. Sant}
\affiliation{Infineon Technologies, Villach, Austria.}
\author{M. Trupke}
\email{michael.trupke@oeaw.ac.at}
\affiliation{Vienna centre for Quantum Science and Technology, Department of Physics, University of Vienna, Boltzmanngasse 5, 1090 Vienna, Austria}
\affiliation{Austrian Academy of Sciences, Institute for Quantum Optics and Quantum Information (IQOQI) Vienna, Boltzmanngasse 3, A-1090 Vienna, Austria}
\date{\today}

\begin{abstract}

Quantum sensors based on the nitrogen-vacancy (NV) centre in diamond are rapidly advancing from scientific exploration towards the first generation of commercial applications. While significant progress has been made in developing suitable methods for the manipulation of the NV centre spin state, the detection of the defect luminescence has so far limited the performance of miniaturized sensor architectures. 
The recent development of photoelectric detection of the NV centre's spin state offers a path to circumvent these limitations, but has to-date required research-grade low current amplifiers to detect the picoampere-scale currents obtained from these systems. Here we report on the photoelectric detection of magnetic resonance (PDMR) with NV ensembles using a complementary metal-oxide semiconductor (CMOS) device. The integrated circuit delivers a digitized output of the diamond sensor with low noise and 50 femtoampere resolution. This integration provides the last missing component on the path to a compact, diamond-based quantum sensor. The device is suited for continuous wave (CW) as well as pulsed operation. We demonstrate its functionality with DC and AC magnetometry up to several megahertz, coherent spin rotation and multi-axial decoupling sequences for quantum sensing.

\end{abstract}

\maketitle

\section{\label{sec:intro}Introduction}

\begin{figure*}[htb]
     \centering
     \includegraphics[width=\textwidth]{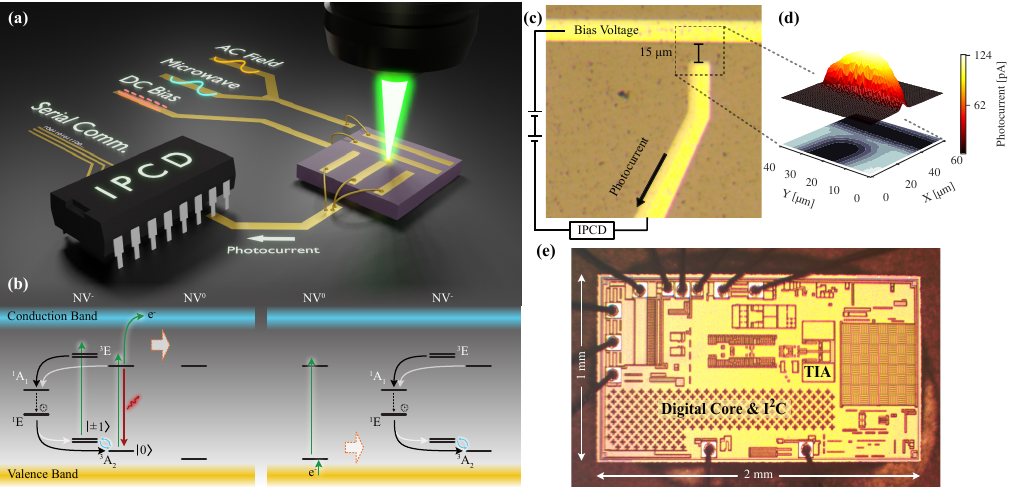}
     \caption{Concept and device. \textbf{(a)} Idealized rendering of the diamond-based sensor chip including the integrated photo-current detector (IPCD). \textbf{(b)} NV centre level diagram and dynamics: The system consists of a triplet ground state $^3A_2$ and the triplet excited state $^3E$, as well as the two singlet levels $^1A_1$ and $^1E$. The latter exhibits a lifetime which is at least one order of magnitude longer than that of all excited states (clock symbol). Without application of an external magnetic field the ground states are split by $D_{gs}$. Transitions can be driven by application of a resonant microwave (MW) signal. Upon excitation, the $m_s=0$ population is likely to yield a fluorescence photon or, under sufficiently strong excitation, result in a two-photon process creating charge carriers collected by the chip. This two photon process ionizes the NV centre to the neutral charge state, from where optical excitation of a valence band electron and recombination at the NV centre can return it to the negative charge state. The $m_s=\pm1$ population tends to vibronically decay to the $^1A_1$ singlet state and will thereafter be shelved for \SI{200} {\nano \second} in the meta-stable state. During this shelving process the system is protected from \SI{520}{\nano \meter} excitation, leading to a reduction of both fluorescence and charge carrier generation from the $m_s=\pm1$ spin states. Following relaxation, the population is distributed among the $m_s=\pm 1$ and $m_s=0$ sub-levels, with a preferential decay into the latter. \textbf{(c)} Microscope picture of the gold structure on the diamond used for photoelectric readout. \textbf{(d)} Measurement scan on the section shown in (c): The lower, 2D contour map shows the fluorescence signal obtained from the diamond. The upper, 3D heat map shows the position-dependent photo-current. 
     \textbf{(e)} Microscope image of the unpackaged IPCD integrated photocurrent detector.}
\label{fig:fig1}
\end{figure*}

Device miniaturization and integration opens up new perspectives for applications of quantum sensors based on atom-like systems in crystals \cite{kim2019,liu2013,patel2016,fedotov2016, chatzidrosos2017,barson2017, webb2019, sturner2019}. Based on its excellent spin coherence time, greater than 1.8 ms for a single NV \cite{balasubramanian2009ultralong}, its optically addressable spin state, and its large dynamic range, the nitrogen-vacancy (NV) centre in diamond \cite{doherty2013} is now firmly established as a room-temperature quantum sensor for crystal stress \cite{grazioso2013measurement}, pressure \cite{doherty2014electronic}, electric fields \cite{dolde2011electric}, temperature \cite{plakhotnik2014all}, and magnetic fields \cite{acosta2010, dolde2011, degen2008, balasubramanian2008, maze2008,rondin2014}, including nuclear magnetic resonance spectroscopy \cite{mamin2013, glenn2018}.

Such systems reach excellent sensitivity, resolution, bandwidth and range, but commonly require large and costly instruments for their manipulation and readout. These factors have so far limited their adoption to highly specialized applications and also hinder the scalability with respect to the number of sensor elements on a single device. Therefore in recent years, significant efforts have been undertaken towards the miniaturization of NV-based sensors \cite{sturner2019, Shalaginov2020,kim2019, Stuerner2021} including their required periphery. Typically, the form factor is strongly affected by the microwave chain, including bulky, high-power amplifiers, and by the optical components required to read out the NV centre's spin state. Regarding spin control, it has been shown that microwave generation can be realized directly on a silicone (Si) chip underneath the diamond \cite{kim2019}. Furthermore, the use of microwave resonators can provide a way to effectively reduce the necessary amplifier size and power at the cost of bandwidth \cite{Bayat2014}. The optical excitation and detection system typically includes a green (510 - \SI{570}{\nano\meter}) excitation laser and single-photon counting modules (SPCM) to collect the red NV centre fluorescence. The excitation apparatus can already be reduced in size by making use of directly driven laser diodes \cite{wang2022portable}, and the integration of the diamond sensor on, or in proximity of, an optical fiber tip has also been explored \cite{fedotov2014fiber,Morley2020, Stuerner2021}.
However, optical detection suffers from the high refractive index of diamond, such that typically only a few percent of the emitted photons actually reach the detectors. This has, to a degree, been circumvented by elaborate surface engineering such as focused ion beam milled solid immersion lenses with NV centres at their centre \cite{marseglia2011,vasconcelos2020} as well as small-scale photodiodes in close proximity to the NV-based sensor \cite{Stuerner2021, sturner2019}. However, all of these systems still lack a compact and robust readout system for the NV centre spin state that can be scaled down to the few- or even single-defect level.\\
The advent of photoelectric detection of the NV centre’s spin state presents a viable path towards integrated electronic detection \cite{bourgeois2015, hrubesch2017, bourgeois2017, gulka2017, siyushev2019, morishita2020room, gulka2021}. Charge carriers are generated at the NV centre in a two-photon ionization process by laser illumination which, in contrast to optical detection, relies on the direct capture of charge carriers through electrodes fabricated onto the diamond surface and does not suffer from total internal reflection due to the relatively high refractive index of diamond. Photoelectric readout furthermore offers the potential to avoid the limiting saturation behavior of optical readout, since the charge carrier generation can occur at a far higher rate than photoemission \cite{bourgeois2020photoelectric}. Chiefly, this method obviates the need for bulky collection optics and mitigates the detrimental effect of the high refractive index on signal acquisition. 

Here we report on the miniaturization and circuit integration of an NV centre based sensor through photocurrent detection with state-of-the-art microchip technology (see Fig. \ref{fig:fig1}a). The small-footprint device features low readout noise, high resolution and a photocurrent signal on the order of \SI{100}{\pico\ampere}, which is detected at room temperature by an integrated photocurrent detector (IPCD). We demonstrate the functionality of the device by detecting both DC and AC magnetic fields, with frequencies larger than one MHz, as well as performing an exemplary decoupling sequence towards advanced quantum sensing applications.
\section{\label{sec:photoSens}Photocurrent detection of the nitrogen-vacancy centre}
The NV centre is formed by replacing two adjacent carbon atoms within the diamond lattice by a nitrogen atom and a vacancy. This configuration forms a stable, atom-sized defect with a triplet ground state. Optically induced spin polarization and spin-dependent fluorescence enable remarkable spin purity and readout contrast, even at room temperature \cite{gruber1997,wirtitsch2023}. The ground state features a zero-field splitting $D_{gs}$ of \SI{2.87}{\giga\hertz} between the $m_s=0$ and $m_s=\pm1$ sub-levels which can be manipulated via magnetic resonance. The resonance frequencies $v_i$ between the $m_s=0$ and $m_s=\pm1$ sub-levels are sensitive to electric and magnetic fields as well as to crystal strain and temperature. In particular, external magnetic fields parallel to the N-V axis induce a Zeeman splitting of the $m_s=\pm 1$ states given by
\begin{equation}
v_i = D_{gs} \pm \gamma B_z,
\end{equation}

where $\gamma$ is the electronic gyromagnetic ratio of \SI{28}{\giga\hertz\per\tesla} and $B_z$ the $z$-component of the magnetic field measured along the NV centre symmetry axis \cite{doherty2013}. Since, due to geometrical constrains within the diamond lattice, NV centres can only be oriented along four axes, ensembles of NV centres can be used to perform vector magnetometry \cite{clevenson2018robust,chen2020vortex,pham2011}.

While optical excitation of the $m_s=0$ sub-level preferentially results in photon emission, the $m_s=\pm1$ populations decay to a meta-stable level with a branching ratio close to 0.5, reducing the fluorescence from these states and allowing spin state readout (see Fig. \ref{fig:fig1}b). Furthermore, the meta-stable level decays to both $m_s=0$ and $m_s=\pm1$, though preferentially towards $m_s=0$, leading to state initialization \cite{doherty2013, wirtitsch2023}.
These mechanisms allow optical detection of magnetic resonances (ODMR), where fluorescence collection upon excitation will decrease if a resonant microwave signal is applied to transfer population from $m_s=0$ to $m_s=1$ or $m_s=-1$.

Similarly to fluorescence readout, in electrical readout the $m_s=0$ state is mostly not shelved into the meta-stable state and can thus produce more charge carriers via two-photon excitation (see Fig. \ref{fig:fig1}b). The excited $m_s=\pm 1$ states instead decay frequently to the long-lived and dark metastable state which has a far smaller ionization cross-section than $^3E$. The basic ODMR method described above can therefore be used without alteration for photoelectric detection of magnetic resonances (PDMR), as shown in Fig. \ref{fig:fig1}b.

\section{On-chip photocurrent detection}
Charge carrier collection is enabled by the application of a bias electric field across the electrode gap, as shown in Fig. \ref{fig:fig1}c and d. We use a voltage of up to \SI{24}{\volt} over a gap of 15 micrometers, produced by a stack of coin cell batteries, resulting in a field gradient of \SI{1.6}{\volt\per\micro\meter}. This value is safely below the limiting value of \SI{3}{\volt\per\micro\meter} given by the breakthrough voltage of air under standard conditions obtained from Paschen's law \cite{Paschen1889}. Cell batteries were chosen for compactness as a low-noise power supply. 
\begin{figure}
     \centering
     \includegraphics[width=0.48\textwidth]{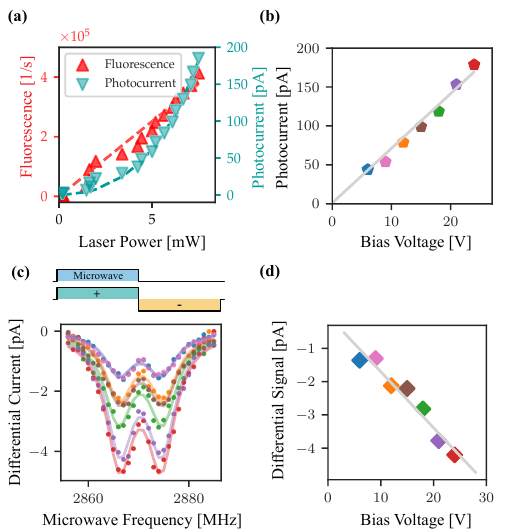}
     \caption{Voltage and Power Dependence. \textbf{(a)} Fluorescence counts (red) and current signal (turquoise) over laser power using a bias voltage of \SI{25}{\volt}. Dashed lines for the fluorescence behaviour are obtained using a linear fit while those for the photocurrent are obtained by fitting $I(P) = \alpha P^2 / (1 + \beta P)$ \cite{siyushev2019} to the data points, where $\alpha \approx 1.26$ and $\beta \approx -0.07$ are constants and $P$ is the applied laser power. \textbf{(b)} Photocurrent versus bias voltage. \textbf{(c)} Microwave frequency sweep using seven different bias voltages. The measurements are performed both with and without application of a microwave pulse to obtain a differential signal. \textbf{(d)} Depth of the dips in (c) versus bias voltage, consistent with a linear dependence of the absolute photocurrent difference.}
\label{fig:fig2}
\end{figure}


The photocurrent is detected by a non-commercial microchip from Infineon Technologies, which acts as an integrated photocurrent detector (IPCD), amplifier, integrator, and digitizer \cite{sant2016}. Its small packaged footprint of only $\SI{1}{\milli\meter}\times \SI{2}{\milli\meter}$ allows to position it in close proximity to the diamond chip (see Fig. \ref{fig:fig1}a for an abstract representation of the setup, and Fig. \ref{fig:fig1}e for a microscope image of the bare chip). The low-noise, ultra-low power (sub-mW) chip offers a 16 bit analog to digital converter with a least-significant-bit (LSB) equivalent current of \SI{50}{\femto\ampere}, a measurement standard deviation of 1.2 LSB, and an integration time of \SI{200}{\milli\second} (see the supplementary information for more details \cite{supplementary}). The chip is powered by a low noise linear regulator at \SI{1.8}{\volt} (ADP150AUJZ-1.8-R7) and addressed as well as read out by a microcontroller, which acts only as an I$^2$C-converter to enable communication with the measurement computer.

Fig. \ref{fig:fig1}d shows a simultaneous fluorescence (bottom) and photocurrent (top) scan over the electrodes where the fluorescence counts and the photocurrent are detected for each position. A clear peak in photocurrent is visible in the middle of the gap between the current detector and bias wires. The fluorescence image shows a clear negative image of the deposited structures.
\begin{figure*}[htb]
     \centering
     \includegraphics[width=\textwidth]{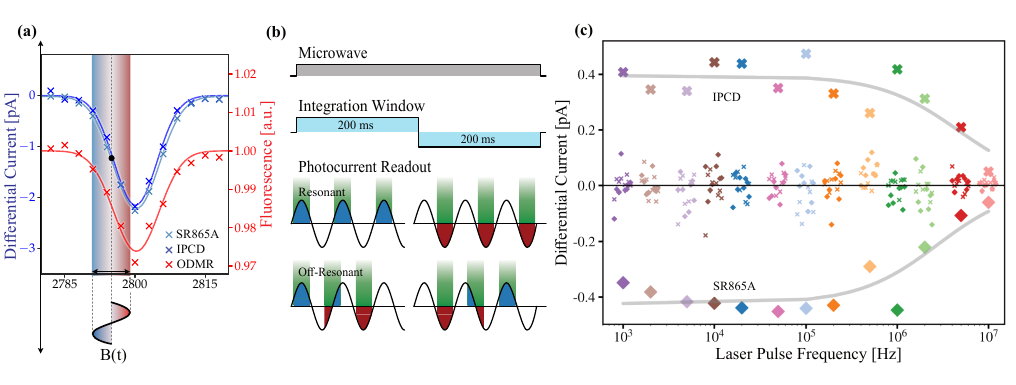}
     \caption{Pulsed-excitation AC Magnetometry. \textbf{(a)} CW optical (red) and photoelectric (blue) detection of magnetic resonance with Gaussian fit functions using a bias voltage of \SI{15}{\volt}. The light blue data points are taken using the Stanford Research SR865A lock-in amplifier while the dark blue lines are obtained using the IPCD. Both traces show the differential signal obtained by performing the measurement with and without microwave. The optically detected trace is normalized to the offset of the Gaussian fit function. The frequency range used for pulsed-excitation AC magnetometry by PSLD measurements is indicated by the vertical shaded bar. Application of a time-dependent magnetic field sensed by the NV ensemble shifts the energy level resonance, leading to a high (blue) or low (red) photocurrent at the field extrema. \textbf{(b)} The measurement is performed in two 200 ms segments phase-matched to the AC-signal. The second segment is phase shifted by an angle of $\pi$ in relation to the first. Within one segment, many green-laser readout pulses induce a photocurrent. In the case where the excitation pulse rate is equal to the modulation rate of the external field, a maximum difference in current is observed between the two segments. In the off-resonant case, the current contributions average to zero for sufficiently large numbers of readout pulses. \textbf{(c)} Application of external AC fields with different frequencies, probed by varying the readout pulse spacing while keeping the pulse width constant. Different colors indicate scans over different external AC fields, e.g. the first (purple) data points show a scan from \SI{0.8}{\kilo \hertz} to \SI{1.2}{\kilo \hertz} with an external AC field of \SI{1}{\kilo \hertz}. We proceed to scan over $\pm 20\%$ of the externally applied field frequencies, which range from \SI{1}{\kilo \hertz} to \SI{10}{\mega \hertz}. We show the differential photocurrent signal using both the IPCD and the commercial lock-in amplifier, with maxima on resonance (large markers) and over a range of four orders of magnitude. The grey lines show a fit to a first order low-pass filter function of the form $I(f) = I_0 / (1 + f/f_0)$, where $I_0$ is the signal amplitude in the low frequency limit, $f_0$ denotes the cut off frequency and $f$ is the frequency. We obtain $f_0 = \SI{5 \pm 2}{\mega\hertz}$ and $f_0 = \SI{3 \pm 1}{\mega\hertz}$ for the cutoff frequencies of the IPCD and the SR865A lock-in measurements respectively. Here, the values for the commercial lock-in amplifier are mirrored around the $x$-axis for better visibility. Small markers indicate measurements where the AC field modulation is out of resonance with the readout pulses.}
\label{fig:fig3}
\end{figure*}

In order to investigate the performance of the IPCD, we perform characterization measurements on the optical and photoelectric characteristics of the diamond sample. First, we perform a saturation scan where we gradually increase the continuous wave (CW) laser power while simultaneously measuring both fluorescence and photocurrent (Fig. \ref{fig:fig2}a). While the power dependence of the fluorescence is consistent with a linear trend, the photocurrent dependence displays the quadratic behaviour expected from the two-photon process involved in charge state conversion (see Fig. \ref{fig:fig1}b). We chose \SI{8}{\milli\watt} excitation power in all subsequent measurements (see SI for further information \cite{supplementary}). Fig. \ref{fig:fig2}b shows the detected photocurrent while varying the bias voltage across the electrode gap. A linear dependence with respect to the applied bias voltage is observed.

We furthermore probe the $m_s=0$ to $m_s=\pm 1$ transitions using an external microwave for seven different bias voltages with no applied magnetic bias field (Fig. \ref{fig:fig2}c) where an intrinsic splitting of \SI{8}{\mega\hertz} between the two spin transitions is discernible. Here, the sequence was divided into two parts: In the first half we include the CW microwave together with CW laser excitation while the second half consists of the readout laser alone. Subtraction of the two sequence segments results in a differential signal which increases linearly with the bias voltage (Fig. \ref{fig:fig2}d).

In a further measurement, the performance of the IPCD is compared to a high-end commercial lock-in amplifier (SRS SR865A). The settings of the commercial device are chosen to match the IPCD parameters as closely as possible. The input impedance is set to \SI{1}{\mega\ohm} and, as in the case of the IPCD measurements, a lock-in frequency of 2 Hz is used, with a time constant of 300 ms and a sensitivity of \SI{5}{\pico\ampere}. We change the bias magnetic field to energetically split the ground state levels equally for all NV centre axis orientations in the ensemble sample to maximize the photoelectric spin signal and allow better distinguishability of resonance shifts under external magnetic fields examined in the next section. The green laser is operated at constant intensity, while the microwave signal is scanned in frequency over the resonance. As above, the photocurrent is extracted in a two-step procedure: a \SI{200}{\milli\second} segment with microwave signal is followed by a second \SI{200}{\milli\second} segment without microwave. The photocurrent readouts of both are subtracted and are shown in Fig. \ref{fig:fig3}a. The measurement shows near-identical results for the IPCD and the commercial lock-in amplifier.

\section{Pulsed-Excitation AC Magnetometry}

To demonstrate the IPCDs capability for AC magnetometry, a sinusoidally-varying current is applied to a wire adjacent to the detection site. The AC signal is generated by a 10 Volt AC source over a \SI{50}{\ohm} dump resistor, resulting in a magnetic field peak-to-peak amplitude of 3.2 G at the readout junction. In order to detect the frequency of this field, we choose the microwave frequency offset for the measurement to coincide with the highest derivative of the resonance signal with respect to magnetic field changes, as indicated by the dotted vertical line in Fig. \ref{fig:fig3}a. A maximum conversion factor of \SI{0.78}{\pico\ampere\per\milli\tesla} is obtained there. The measurement sequence is illustrated in Fig. \ref{fig:fig3}b. In order to detect periodic magnetic field variations with periods much shorter than the photocurrent integration window, we implement a method which we term pulsed-laser stroboscopic detection (PLSD) method: The excitation laser is modulated with square-wave pulses at the target detection frequency. We adjust the pulse duration $\tau_L$ to cover 1/4 of the target AC field period.  PLSD operates in a phase-locked manner, where the readout sequence is again split into two \SI{200}{\milli\second} segments. The optical pulses for photocurrent generation are shifted by a phase of $\pi$ in the second segment. If the readout is performed resonantly with the AC-field oscillation, the current difference between the two segments will be maximal, while off resonance (and with many repetitions of readout pulses per segment) the obtained currents in the segments average out to zero. 
Similarly to quantum heterodyne detection or coherently averaged synchronized readout, the frequency resolution of PLSD is not limited by the coherence lifetime of the NV centres \cite{schmitt2017,glenn2018}. We further note that, by including all four quadratures instead of only two, it is possible to measure AC fields of arbitrary phase with this approach, at the cost of doubling the readout duration.

The results of the PLSD measurements are shown in Fig. \ref{fig:fig3}c. We sweep the laser pulse frequency to probe several external AC fields and keep the laser pulse width constant while varying the inter-pulse distance (see Fig. \ref{fig:fig3}b). We obtain clear signal peaks whenever the pulsed laser readout has the same frequency as the external field. Here, the current values for the SR865A are mirrored around the $x$-axis for better visibility. The two devices measure near-identical currents on resonance (large markers in the figure) while showing no significant current when detuned (small symbols). The measurement covers over four orders of magnitude in frequency and demonstrates that PLSD allows to measure AC fields with a \SI{3}{\decibel} bandwidth of \SI{3}{\mega\hertz}.

\begin{figure*}
     \centering
     \includegraphics[width=0.6\textwidth]{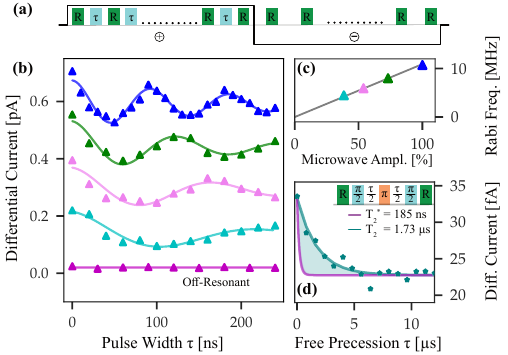}
     \caption{Coherent spin rotation. \textbf{(a)} Sequence used for photoelectric detection of NV Rabi oscillations. During the first half we interleave \SI{5}{\micro\second} long laser pulses (R) with \SI{8}{\milli\watt} and microwave pulses with a variable pulse width ($\tau$). We omit the microwave pulses during the second half to obtain a differential signal. \textbf{(b)} Rabi oscillations for several microwave powers, in agreement with the expected linear increase in oscillation frequency with increased microwave power. The curves are shifted vertically by \SI{0.15}{\pico\ampere} increments for clarity. We furthermore perform an additional measurement (magenta), far off resonance from the $m_s=0$ to $m_s=1$ transition, to exclude parasitic effects such as cross-talk from the high-power microwave line in close proximity to the readout chip. \textbf{(c)} Extracted Rabi frequencies from (b) show a linear dependence on the microwave amplitude. \textbf{(d)} CPMG sequence to determine the $T_2$ coherence time of the NV ensemble. This sequence typically forms the basis for more elaborate sequences used in quantum sensing experiments. First, a $\pi / 2$ pulse around the $x$-axis of the Bloch sphere rotates the spin into a superposition of $m_s=0$ and $m_s= 1$, followed by a free precession time $\tau/2$. Dephasing induced by slowly varying magnetic fields is then cancelled by the application of a $\pi$ rotation around the $y$-axis, followed by an additional free precession time $\tau/2$ for rephasing. The magenta line shows the (shorter) $T_2^\star$ coherence time of $\sim$ \SI{185}{\nano\second} extracted from the decay envelope of the Rabi oscillations in (b). Teal data points show the prolonged $T_2$ coherence time, while the solid line is an exponential fit to the data. We extract a coherence time of $T_2 = \SI{1.73}{\micro\second}$.}
\label{fig:fig4}
\end{figure*}

\section{Coherent spin manipulation}
While so far, the microwave signal was operated in a continuous fashion and the laser was pulsed for AC magnetometry, many sensing applications using the NV centre strongly benefit from the application of pulsed spin rotation sequences \cite{rondin2014,grotz2011,schmitt2017}. We thus show that the integrated chip is suitable for pulsed-sequence readout by detecting coherent rotations of the spins. 

To mitigate inhomogeneous driving and remove the spectral overlap of the four NV orientations in the high-density sample, we first single out one NV orientation using a neodymium magnet to apply a bias field along one of the crystalline orientations. Using the same measurement envelope as in Fig. \ref{fig:fig2}c, we alternate laser and microwave pulses while varying the microwave pulse length $\tau$ (see Fig. \ref{fig:fig4}a). Fig. \ref{fig:fig4}b shows Rabi oscillations with four different microwave powers. We obtain the expected sinusoidal signal (solid lines) with an additional damping due to the $T_2^\star$ coherence time of the sample. Furthermore, the expected linear relationship between microwave amplitude and Rabi frequency is visible in Fig \ref{fig:fig4}c. We additionally perform a measurement under full microwave power, off resonance from the $m_s=0$ to $m_s=1$ transition, where no change in photocurrent versus pulse width is observed. This measurement shows that there is negligible cross-talk between the high-power microwave line and the IPCD (darker pink line and data points in Fig. \ref{fig:fig4}b). 

The most advanced quantum sensing methods make use of multi-axial decoupling methods to minimize the effects of stray magnetic fields and control pulse imperfections \cite{Barry2020}. Fig. \ref{fig:fig4}d shows the results when applying a Carr-Purcell-Meiboom-Gill (CPMG) sequence in order to determine the $T_2$ coherence time of the sample. Compared to the CW-microwave, pulsed-laser method shown in Fig. \ref{fig:fig3}, this method enables to sense external fields at arbitrary frequencies, ranging from DC to an upper limit given by the hardware limit determined only by the attainable Rabi frequency.
First, a $\pi/2$ pulse about the $x$-axis of the Bloch sphere rotates the initial state into a coherent superposition of $m_s=0$ and $m_s= 1$. During a subsequent free precession time of length $\tau/2$, the superposition will rotate unpredictably along the equator of the Bloch sphere due to magnetic field noise from adjacent spins and external sources. Those external disturbances lead to the comparatively short $T_2^\star$ coherence time extracted from the decay envelope of the Rabi oscillations in Fig. \ref{fig:fig4}b. However, subsequent application of a $\pi$ pulse about the orthogonal $y-$axis, followed by an additional free precession time $\tau/2$ allows the reversal of a large part of these dephasing mechanisms. The sequence is concluded by a $\pi/2$ pulse which, were it not for irreversible decoherence processes, would lead to a near-complete spin state transfer from $m_s=0$ to $m_s = 1$. The pulses are generated by a field programmable gate array (FPGA) \cite{reisenbauer2022}. We obtain a $T_2$ coherence time of $\sim\, \SI{1.73}{\micro\second}$, corresponding to an improvement by more than an order of magnitude compared to the bare dephasing time $T_2^\star$. This sequence can already be used to sense external AC magnetic fields with high sensitivity \cite{rondin2014}. 
 
 We note here that lock-in based methods such as photoelectric detection presently integrate over the entire laser pulse duration, while optical detection allows to select the relevant time window of $\sim$ \SI{300}{\nano\second} at the beginning of the readout pulse, allowing to optimize the optical readout contrast \cite{magaletti2024,wirtitsch2023}. Fast gating methods for the electrical readout current are therefore of great interest for further improvement of the technique.

\section{Discussion and Conclusion}
This work combines several milestones for diamond quantum sensors on the path from the laboratory towards practical applications. Together with previous demonstrations, including small-scale optical excitation \cite{sturner2019} and chip-integrated microwave generation \cite{kim2019,Shalaginov2020} to manipulate the spin state, the integration of electrical readout of the NV centre spin state shown in this work completes the foundation for extremely compact, low-power quantum sensors. The implemented device is CMOS compatible and enables photoelectric readout of nitrogen vacancy centres in diamond with sufficiently low noise and high current resolution to even enable the readout of single centres \cite{siyushev2019}. Due to its low noise and large dynamic range, the IPCD can be used to read out both continuous-wave and pulsed sequences, frequently encountered in quantum sensing applications. The IPCD achieves near-identical performance as a state-of-the-art lock-in amplifier, demonstrating both DC as well as AC magnetometry. The DC sensitivity of \SI{1.6}{\micro\tesla\per\sqrt{\hertz}} reached (see methods) in this first demonstration already compares favorably with previous approaches \cite{kim2019}, while necessitating far lower excitation power, and can be improved dramatically with optimized substrates \cite{bourgeois2020photoelectric}.
Furthermore, the first stepping stones towards the use of more complex dynamical decoupling schemes are shown, underscoring the functionality of the device for a broad range of diamond quantum sensor applications. 

The combination with board- or chip-level integration of optical excitation, microwave generation, and sequence control will complete the development towards miniaturized, high-performance diamond quantum sensors based on photoelectric readout.

\section{\label{sec:Methods} Methods}
\subsection{\label{sec:chip} Fabrication and Setup}
A high-pressure, high-temperature (HPHT) diamond from Element-6 with a nitrogen concentration $<$200 ppm served as the NV centre host. It was electron irradiated at 100°C and annealed at 900°C to increase NV centre yield and resulting in an NV concentration of $8\pm2\,$ppm. After cleaning in a mixture of potassium nitrate and sulfuric acid, followed by repeated cleaning in solvents (acetone, isopropanol, ethanol, and methanol) the sample was further left in oxygen plasma for five minutes at 100 W. Subsequently, a custom-made silicon hard mask was positioned on it, immediately followed by another oxygen plasma cleaning step and metal deposition. First, a \SI{100}{\nano\meter} layer of titanium was sputtered onto the sample, followed by \SI{100}{\nano\meter} of thermally evaporated copper and \SI{20}{\nano\meter} of gold to avoid oxidation of the copper layer. The electrode arrangement is shown in Fig. \ref{fig:fig1}c. The width of the wires and the gap size are both \SI{15}{\micro\meter}.

After deposition, the sample was repeatedly cleaned in the previously mentioned solvents, mounted and bonded onto a printed circuit board (PCB), with electrical connections formed by \SI{20}{\micro\meter} thick aluminium bond wires (details in supplementary information \cite{supplementary}). The PCB provides an electrical interface to our measurement and manipulation hardware, as well as to the commercial lock-in amplifier for device comparison \cite{supplementary}. While currently a microcontroller reads out each part of the differential signal and the values are subtracted in the measurement computer, the IPCD can be altered to provide the  background subtracted signal directly, obviating the need for the microcontroller. This direct lock-in mode is already available, albeit with an unsuitable combination of lock-in frequency and signal gain. Furthermore, a built-in feature of the IPCD is given by a synchronous TTL output, in phase with the lock-in cycle, which can be used to directly control the laser driver or microwave switch such that no external TTL source would be required.

In order to compare photoelectric readout to conventional, optically detected magnetic resonance, the PCB was mounted into a confocal microscope setup using an Olympus microscope objective with a numerical aperture of 0.85. A laser diode provided green \SI{520}{\nano\meter} excitation, necessary for spin initialization and readout, with a maximum power of \SI{9}{\milli\watt}. The photoluminescence was filtered with a $>$\SI{650}{\nano\meter} longpass filter. We underline that, while collection path filtering and photodetection are not needed for electrical readout, an excitation path which provides a sufficiently high intensity (on the order of \SI{50}{\milli\watt\per\micro\meter\squared}) remains necessary.

The microwave signal necessary to drive the spin transitions was generated by an Analog Devices ADF4351 RF synthesizer, switched using a Mini-Circuits ZASWA2-50-DR+ RF switch, and amplified by \SI{45}{\decibel} with a Mini-Circuits ZHL-16W-43+ high power amplifier. The amplified signal was combined with an AC source (0-500 MHz) used for the PLSD measurements using a custom made diplexer (Wainwright Instruments). The combined signal output was connected to the PCB, reaching the the microfabricated wire on the diamond chip through a bond wire.

\subsection{\label{sec:ccomparison} Comparison of Photoelectric and Optical Readout}
In photoluminescence readout, we observe a photon collection rate of $R=4\times10^5$ cts/s from the NV ensemble after reducing the collection efficiency with neutral density (ND) filters by 5.7 orders of magnitude, and including all system losses. Removing the ND filters  would result in a rate of $R=2.2\times10^{11}$ cts/s. The minimum detectable field in a CW ODMR magnetometry measurement for small fields is given by \cite{rondin2014}
\begin{equation}
B_{\text{min}}^{\text{(DC)}} = \frac{\sqrt{2}}{\cos{\theta}}\frac{4}{3\sqrt{3}}\frac{h}{g_e\mu_B}\frac{\Delta\nu}{C_{\text{CW}}\sqrt{R}}\frac{1}{\sqrt{\tau_{\text{int}}}},
\label{eqn:sensCW}
\end{equation}
where $\Delta\nu$ is the full width at half-maximum of the lineshape, $h$ the Planck constant, g the Land\'e factor, $\mu_B$ the Bohr magneton, $C_{CW}$ the readout contrast and $R$ the average detection rate. The angle $\theta$ between the NV centre axis and the applied field leads to an increase of $\sqrt{3}$. The $\sqrt{2}$ factor accounts for the equal bisection of the sequence into a measurement part, with the microwave field applied, and a normalization measurement without. Using the values from the measurement in Fig. \ref{fig:fig3} ($\Delta\nu=\SI{11}{\mega\hertz}$, $C_{\text{CW}}=2.6\,$\%), we find a raw sensitivity for optical detection of \SI{53.2}{\micro\tesla\per\sqrt{\hertz}}, while correcting for the attenuation by the ND filters yields a nominal sensitivity of \SI{71}{\nano\tesla\per\sqrt{\hertz}}. The nominal sensitivity for electrical readout results instead from the rate of charge carriers collected in the system. There, with an average photocurrent of \SI{75}{\pico\ampere}, the rate is $4.7\times10^8\,$s$^{-1}$ yielding a sensitivity of \SI{1.6}{\micro\tesla\per\sqrt{\hertz}}.

The sensitivity of the PLSD measurements at a frequency $f$ is reduced by the duty cycle of $f\tau_L$, with $\tau_L$ the duration of the illumination pulse. Additionally, the average detected field during a readout pulse is reduced, compared to a DC field, because of the sinusoidal time dependence. However, both portions of the sequence are used to measure the magnetic field, improving the sensitivity by $\sqrt{2}$. These factors lead to a minimum detectable field of
\begin{equation}
B_{\text{min}}^{\text{(PLSD)}}=B_{\text{min}}^{\text{(DC)}}\frac{\pi \sqrt{f\tau_L}}{\sqrt{2}\sin{(\pi f\tau_L)}}.
\label{eqn:sensPLSD}
\end{equation}
We use a duty cycle of $25\,$\%, resulting in a nominal sensitivity of \SI{2.4}{\micro\tesla\per\sqrt{\hertz}}.

All sensitivity values achieved here can be improved significantly by optimizing the material properties of the diamond. The defect density is a key parameter: For decreasing nitrogen and NV centre density, the spin resonance linewidth decreases and the photocurrent per NV centre increases, while the photoluminescence per NV centre remains constant. It will therefore be possible to generate similar photocurrents while drastically reducing the spin resonance linewidth, thus improving the sensitivity of the device, by using tailored diamond substrates. Furthermore, recent experiments under similar conditions on a different diamond have resulted in a photocurrent of \SI{4}{\nano\ampere} at similar laser powers, translating to a sensitivity improvement by an order of magnitude \cite{bourgeois2020photoelectric}. 

Even better sensitivity is possible using a larger detection volume and higher excitation power \cite{villaret2023}. From the obtained photon count rate and estimated NV centre density in our sample we calculate the sensor volume using our excitation beam to be \SI{3.2}{\micro\meter\cubed}. However, we note that the volume extracted from photon collection will most likely be larger than the actual interrogation volume available for photocurrent detection, due to the latter's quadratic dependence on laser power. Nonetheless, using this conservative estimate, fabrication of interdigitated electrodes across the entire diamond surface would allow to improve the sensitivity significantly: For a \SI{3}{\milli\meter}$\times$\SI{3}{\milli\meter}$\times$\SI{10}{\micro\meter} interrogation volume, the nominal sensitivity would improve by another 5 orders of magnitude, reaching $\SI{33}{\pico \tesla \per \sqrt{\hertz}}$. Further improvement towards \SI{1}{\pico\tesla} and below will be possible using NV ensembles with narrower spin resonance linewidth and optimized photocurrent collection.

\subsection{\label{sec:noise} Noise Estimation}
The IPCD exhibits a noise floor of 1.2 least-significant bits (LSB), corresponding to a nominal noise floor of \SI{84}{\femto\ampere\per\sqrt{Hz}} which represents the largest of the contributing noise factors. The electron shot noise at \SI{75}{\pico\ampere} is \SI{4.8}{\femto\ampere\per\sqrt{Hz}}, an order of magnitude smaller than the bit noise, and the thermal Johnson-Nyquist noise is \SI{0.6}{\femto\ampere\per\sqrt{Hz}}. Inherent noise factors are thus still almost two orders of magnitude below our detected differential signal of \SI{2}{\pico\ampere} in CW-PDMR. The relative noise floor can be reduced by increasing the photoelectric signal, either by use of a different diamond host that produces a larger photoelectric current, or by implementing low-noise, chip-scale pre-amplification of the obtained signal\cite{djekic2017transimpedance}.

\section{Acknowledgments}
We wish to acknowledge the support from the FFG via projects 870002 (QSense4Life) and 877615 (QSense4Power), and the QuantERA FFG project 864036 (Q-Magine). This project has received funding from the European Union’s Horizon Europe research and innovation program under Grant Agreement No. 101046911 (QuMicro). We thank J. Pribo\v{s}ek, K. Harms, T. Moldaschl, G. Aub\"ock, C. Hirschl, F. Starmans, and F. Michl for useful discussions.

\textbf{Conflict of Interest:} The authors declare no conflict of interest.
\appendix


\bibliography{aipsamp}

\end{document}